\RequirePackage{arydshln}
\documentclass[aps,twocolumn,nofootinbib,superscriptaddress,preprintnumbers,pra,10pt,floatfix]{revtex4-1}

\usepackage[utf8]{inputenc}

\usepackage{float}
\usepackage{amsmath,amssymb}
\usepackage{dsfont} 
\usepackage{hyperref}
\usepackage{graphicx}
\usepackage{enumitem}
\usepackage{mathtools}
\usepackage{bbold}
\usepackage{multirow}
\usepackage{ytableau}
\usepackage{youngtab}
\usepackage{braket}
\usepackage{soul}
\usepackage{cancel}
\usepackage{xcolor}
\usepackage[normalem]{ulem}

\usepackage{lipsum}

\usepackage{colortbl} % Always load before 'arydshln'. Otherwise, vertical dashed lines in tables disappear
\definecolor{Gray}{gray}{0.95}
\definecolor{RGray}{gray}{0.90}
\definecolor{CGray}{gray}{0.92}

\usepackage{arydshln}

\usepackage{tikz}
\usetikzlibrary{calc,tikzmark,fit,shapes.geometric,matrix,decorations.markings,arrows.meta,decorations.pathmorphing,patterns,positioning,snakes}

\tikzset
  {midarrow/.style={decoration={markings,mark=at position 0.5 with
     {\arrow[thin,xshift=2pt]{Triangle[length=4pt,#1]}}},postaction={decorate}}
  }

\tikzset{
proton/.style = {circle, draw=black, thin, fill=black!20!white, minimum size=#1,
              inner sep=0pt, outer sep=0pt},
proton/.default = 6pt % size of the circle diameter 
}

\tikzset{
blob/.style = {circle, draw=black, thin, preaction={fill, black!20!white}, pattern=north east lines, minimum size=#1,
              inner sep=0pt, outer sep=0pt},
blob/.default = 6pt % size of the circle diameter 
}

\tikzset{
wc/.style = {circle, fill, minimum size=#1,
              inner sep=0pt, outer sep=0pt},
wc/.default = 4pt % size of the circle diameter 
}

\tikzset{vector/.style={decorate, decoration=snake}}

\makeatletter
\g@addto@macro\bfseries{\boldmath}
\makeatother

\makeatletter
\renewcommand\paragraph{\@startsection{paragraph}{4}{\z@}%
                                    {3.25ex \@plus1ex \@minus.2ex}%
                                    {-1em}%
                                    {\normalfont\normalsize\bfseries}}
\makeatother

\allowdisplaybreaks
\begin{document}

\preprint{}
\preprint{}

\title{Understanding the first measurement of $\mathcal{B}(B\to K \nu \bar{\nu})$}

\author{L.~Allwicher}
\email{lukall@physik.uzh.ch}
\affiliation{Physik-Institut, Universit\"{a}t Z\"{u}rich, CH-8057 Z\"{u}rich, Switzerland}
\author{D.~Be\v{c}irevi\'c}
\email{damir.becirevic@ijclab.in2p3.fr}
\affiliation{IJCLab, P\^ole Th\'eorie (Bat.~210), CNRS/IN2P3 et Universit\'e, Paris-Saclay, 91405 Orsay, France}
\author{G.~Piazza}
\email{gioacchino.piazza@ijclab.in2p3.fr}
\affiliation{IJCLab, P\^ole Th\'eorie (Bat.~210), CNRS/IN2P3 et Universit\'e, Paris-Saclay, 91405 Orsay, France}
\author{S.~Rosauro-Alcaraz}
\email{salvador.rosauro@ijclab.in2p3.fr}
\affiliation{IJCLab, P\^ole Th\'eorie (Bat.~210), CNRS/IN2P3 et Universit\'e, Paris-Saclay, 91405 Orsay, France}
\author{O.~Sumensari}
\email{olcyr.sumensari@ijclab.in2p3.fr}
\affiliation{IJCLab, P\^ole Th\'eorie (Bat.~210), CNRS/IN2P3 et Universit\'e, Paris-Saclay, 91405 Orsay, France}

\begin{abstract}
\vspace{5mm}
Recently, Belle~II reported on the first measurement of $\mathcal{B}(B^\pm\to K^\pm \nu\bar{\nu})$ which appears to be almost $3\sigma$ larger than 
predicted in the Standard Model. We point out the important correlation with 
$\mathcal{B}(B\to K^{\ast} \nu\bar{\nu})$ so that the measurement of that decay mode could help restrain the possible options for building the model of New Physics. We interpret this new experimental result in terms of physics beyond the Standard Model by using SMEFT and find that a scenario with coupling only 
to $\tau$ can accommodate the current experimental constraints but fails in getting a desired $R_{D^{(\ast )}}^\mathrm{exp}/R_{D^{(\ast )}}^\mathrm{SM}$, unless one turns the other SMEFT operators that are not related to $b\to s\ell\ell$ or/and $b\to s\nu\nu$. 
\vspace{3mm}
\end{abstract}

\maketitle

\allowdisplaybreaks

%%%%%%%%%%%%%%%%%%%%%%%%%%%%%%%%%%%%%%%%
\section{Introduction}\label{sec:intro}
%%%%%%%%%%%%%%%%%%%%%%%%%%%%%%%%%%%%%%%%

Experimental studies of the angular distribution of $B\to K^\ast (\to K\pi) \mu\mu$~\cite{LHCb:2015svh,CMS:2020oqb} and $B\to K \mu\mu$~\cite{LHCb:2014auh,CMS:2018qih} offered a number of observables, the study of which could help unveiling the effects of presence of New Physics (NP), i.e. physics beyond the Standard Model (BSM). It appeared, however, that interpreting several apparent deviations with respect to the Standard Model (SM) predictions required a very good control over the hadronic uncertainties, and in particular those related to the low-energy operators coupling to $\bar c \gamma_\mu c$. Those latter contributions are particularly problematic in the regions of $q^2 =(p_{\mu^+}+p_{\mu^-})^2$ populated by the $c\bar c$ resonances. Clearly, in order to resolve the effects of BSM physics from those related to the SM weak interaction in these regions, one needs to evaluate the relevant hadronic matrix element of a non-local operator, which cannot be done yet in lattice QCD. Instead, one resorts to using various assumptions such as the quark-hadron duality to treat the problem perturbatively (even though the energy window is narrow)~\cite{Greub:2008cy,Asatryan:2001zw,Asatrian:2019kbk}, or a specific hadronic model~\cite{Khodjamirian:2010vf,Khodjamirian:2012rm,Gubernari:2020eft,Gubernari:2022hxn}. That problem is commonly circumvented when measuring $\smash{R_{K^{(\ast)}} = \mathcal{B}^\prime(B\to K^{(\ast )} \mu\mu)/ \mathcal{B}^\prime(B\to K^{(\ast )} e e )}$~\cite{Hiller:2003js}, where $\mathcal{B}^\prime$ is used to indicate that the partial branching fractions are measured in the interval $q^2\in [1.1, 6]~\mathrm{GeV}^2$, well below the first $c\bar c$ resonance, $m_{J/\psi}^2 = (3.097\,\mathrm{GeV})^2$. First measurements of $R_K$ and $R_{K^\ast}$, as well as of $\mathcal{B}(B_s\to \mu\mu)$~\cite{ATLAS:2018cur,CMS:2020rox,LHCb:2021awg,CMS:2022mgd} indicated an important departure from the SM predictions. Many models used to accommodate these deviations were used to constrain the BSM couplings, most of which also implied a significant deviation of $\mathcal{B}(B\to K^{(\ast )} \nu\bar\nu)$ from its SM value. 
Very recently, however, LHCb reported $R_K=0.998(90)$, $R_{K^\ast}=0.930(97)$~\cite{LHCb:2022qnv}, thus fully consistent with $R_{K^{(\ast )}}^\mathrm{SM}=1.00(1)$~\cite{Bordone:2016gaq}.

Even though it is experimentally much more challenging, $B\to K^{(\ast )} \nu\bar\nu$ is theoretically cleaner than the equivalent mode with charged leptons instead of neutrinos in the final state. This is so because the coupling to the problematic operators involving $c\bar c$ resonances is absent. The remaining non-perturbative QCD obstacle, that both decay modes share, is a reliable theoretical estimate of the hadronic matrix elements of the local operators in the entire physical region, $0 \lesssim q^2\leq (m_B-m_{K^{(\ast )}} )^2$. That task is also very challenging for lattice QCD because the available $q^2$ range is too large. This is why the lattice QCD results are used to constrain the parameters entering a model $q^2$-dependence of the relevant form factor, necessary for the SM prediction of $\mathcal{B}(B\to K \nu\bar\nu)$. Interestingly, in the case of $B\to K \nu\bar\nu$ the $q^2$ shape of the hadronic form factor can be checked experimentally by measuring the partial branching fractions $\mathcal{B}^\prime (B\to K \nu\bar\nu)$, as discussed in Ref.~\cite{Becirevic:2023aov}.

In the following we will often use the ratio $R_{\nu\nu}^{K^{(\ast )}}= \mathcal{B}(B\to K^{(\ast )} \nu\bar\nu)/\mathcal{B}(B\to K^{(\ast )} \nu\bar\nu)^\mathrm{SM}$, for which the experimental upper bounds exist ~\cite{Belle:2017oht}: 
 \begin{align}
  &\label{eq:belleBounds1}
  R^{K}_{\nu\nu} < 3.6 \quad(90\% \ \text{C.L.})\,,\\
 &\label{eq:belleBounds2}
 R^{K^\ast}_{\nu\nu} < 2.7 \quad(90\% \ \text{C.L.})\,.
\end{align}
Very recently, the first of these bounds was superseded by the first measurement of this decay mode by Belle~II, $\mathcal{B}(B^\pm\to K^\pm \nu\bar\nu) =  2.40(67)\times 10^{-5}$~\cite{Belle-II:2023esi}, which appears to be 
$2.9\sigma$ larger than its SM estimate.~\footnote{Note that for the SM estimate we take  $\mathcal{B}(B^\pm\to K^\pm \nu\bar\nu) = 4.44(14)(27) \times 10^{-5}$\cite{Becirevic:2023aov} i.e.~the value which does not include the tree level contribution~\cite{Kamenik:2009kc} which was also subtracted away during the experimental data analysis~\cite{Belle-II:2023esi}.}
This then leads to  
 \begin{align}
 R^{K}_{\nu\nu} =5.4\pm 1.5\,.
\end{align}
In this letter we discuss how this departure from the SM prediction can be interpreted in terms of generic BSM scenarios in the SMEFT framework, which then requires one to remain consistent with the stringent bounds on NP in $b\to s$ transitions arising from the measured $R_{K^{(\ast)}}$ mentioned above, as well as from the measured $\mathcal{B}(B_s\to \mu\mu)= 3.35(27) \times 10^{-8}$~\cite{CMS:2022mgd}, which is consistent with the SM estimate $\mathcal{B}(B_s\to \mu\mu)^\mathrm{SM}= 3.66(4) \times 10^{-8}$~\cite{Beneke:2019slt}.

\section{Effective Theory considerations}

\begin{figure}[t!]
\centering
\centerline{ \includegraphics[width=1.35\linewidth]{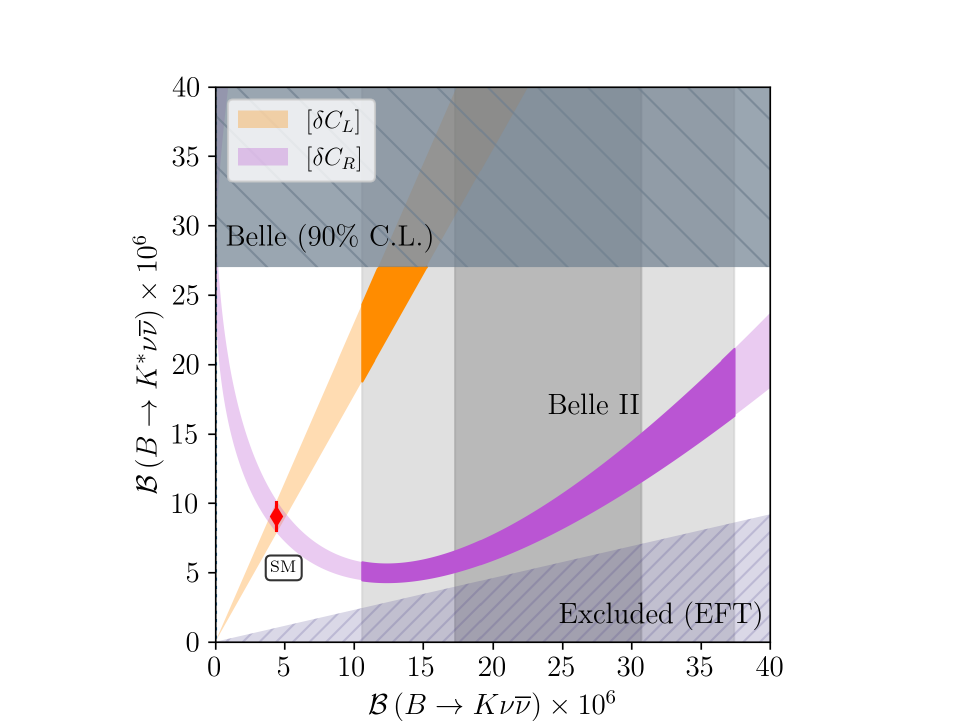}}
\caption{\small \sl
The correlation between $\mathcal{B}(B\to {K^{\ast}}{\nu\nu})$ and $\mathcal{B}(B\to K{\nu\nu})$ decays with respect to the variation of 
$\delta C_{L}$ or $\delta C_{R}$. The shaded gray area correspond to $1\sigma$ and $2\sigma$ of the recent Belle~II result for $\mathcal{B}(B\to K{\nu\nu})$. The red point corresponds to the SM predictions for these observables. We also show the region of experimentally excluded $\mathcal{B}(B\to {K^{\ast}}{\nu\nu})$ values~\cite{Belle:2017oht} (gray hatched area), as well as the region which is not accessible within the EFT approach  (purple hatched area), cf.~Eq.~\eqref{eq:eft-prediction}}. 
\label{fig:corr1} 
\end{figure}

The low energy effective theory relevant to the $b\to s\nu\nu$ decay is described by the effective Hamiltonian, 
%%%%%%%%%%%%%%
\begin{align}
\label{eq:eft-bsnunu}
\mathcal{L}_\mathrm{eff}^{\mathrm{b\to s\nu\nu}} =  \dfrac{4 G_F}{\sqrt{2}} \lambda_t \sum_a C_a\, \mathcal{O}_a+\mathrm{h.c.}\,, 
\end{align}
%%%%%%%%%%%%%
with
\begin{align}
\label{eq:eft-ops}
\mathcal{O}_{L}^{\nu_i\nu_j} &=\dfrac{e^2}{(4\pi)^2}(\bar{s}_L \gamma_\mu b_L)(\bar{\nu}_i \gamma^\mu (1-\gamma_5)\nu_j)\,,\nonumber\\[0.3em]
\mathcal{O}_R^{\nu_i\nu_j} &=\dfrac{e^2}{(4\pi)^2}(\bar{s}_R \gamma_\mu b_R)(\bar{\nu}_i \gamma^\mu (1-\gamma_5)\nu_j)\,,
\end{align}
in a standard notation with $\lambda_t = V_{tb} V_{ts}^\ast$, the suitable combination of the Cabibbo–Kobayashi–Maskawa (CKM) matrix elements. In the SM we know that $C_R^{\mathrm{SM}}=0$ and $\big{[}C_{L}^{\nu_i\nu_j}\big{]}_\mathrm{SM}\equiv\delta_{ij}\, C_{L}^{\mathrm{SM}}$, with    
$C_L^\mathrm{SM}= -6.32(7)$~\cite{Buchalla:1993bv,Buchalla:1998ba,Misiak:1999yg,Brod:2010hi,Buras:2014fpa}.
A detailed discussion of hadronic uncertainties and those arising from the choice of $\lambda_t$ can be found in Ref.~\cite{Becirevic:2023aov}. 
Here we will just mention that in the SM the values of decay rates of the charged $B$-meson read:
\begin{align}\label{eq:SMnunu}
\mathcal{B}(B^\pm\to K^\pm\nu\nu) &= (4.44\pm 0.30) \times 10^{-6}\,,\nonumber\\[0.3em]
\mathcal{B}(B^\pm\to K^{\pm\ast}\nu\nu) &= (9.8\pm 1.4) \times 10^{-6}\,,
\end{align}
and they do not comprise the triply Cabibbo-suppressed tree-level contribution, $B^{+}\rightarrow \tau^{+}\left(\rightarrow K^+\bar{\nu}\right)\nu$ (also subtracted away in the experimental data analysis~\cite{Belle-II:2023esi}). 
It is convenient to factor out the SM contribution to the branching fractions and write
\begin{align}
\mathcal{B}(B\to {K^{(\ast)}}{\nu\nu}) &= \mathcal{B}(B\to {K^{(\ast)}}{\nu\nu})\Big{|}_\mathrm{SM} \, \big{(}1+\delta \mathcal{B}_{K^{(\ast)}}^{\nu\nu}\big{)}\,,
\end{align}
%%%%%%%%%%%%%
so that the NP piece, $\delta \mathcal{B}_{K^{(\ast)}}^{\nu\bar{\nu}}$, can be expressed in terms of $\delta C_{L,R}^{\nu_i\nu_j}$, the BSM contribution to the left- or right-handed operator given in Eq.~\eqref{eq:eft-ops}. It is then straightforward to express $R_{\nu\nu}^{K^{(\ast)}}=1+\delta\mathcal{B}_{K^{(\ast)}}^{\nu\bar{\nu}}$. If we write $C_{L,R}^{\nu_i\nu_j}=  \delta_{ij} C_{L,R}^\mathrm{SM} + \delta C_{L,R}^{\nu_i\nu_j}$, then we have~\cite{Buras:2014fpa,Becirevic:2023aov}: 
%%%%%%%%%%%%
\begin{align}
\begin{split}
\delta \mathcal{B}_{K^{(\ast)}}^{\nu\bar{\nu}} &=  \sum_{i}\dfrac{2\mathrm{Re}[C_L^\mathrm{SM}\,(\delta C_{L}^{\nu_i\nu_i}+\delta C_{R}^{\nu_i\nu_i})]}{3|C_{L}^\mathrm{SM}|^2}\\
&+\sum_{i,j}\dfrac{|\delta C_{L}^{\nu_i\nu_j}+\delta C_{R}^{\nu_i\nu_j}|^2}{3|C_L^\mathrm{SM}|^2}\\
&- \eta_{K^{(\ast)}}\sum_{i,j} \dfrac{\mathrm{Re}[\delta C_R^{\nu_i\nu_j}(C_{L}^\mathrm{SM}\delta_{ij}+\delta C_{L}^{\nu_i\nu_j})]}{3|C_{L}^\mathrm{SM}|^2}\,,
\end{split}
\end{align}
%%%%%%%%%%%%
\noindent where the sum over neutrino flavor indices is understood, $i,j \in \lbrace 1,2,3 \rbrace$. In the above expression, $\eta_K=0$, and $\eta_{K^\ast}=3.33(7)$. 
In that way, %and in terms of $\delta C_{L}$
 one can easily check the response of $\mathcal{B}(B\to {K^{\ast}}{\nu\nu})$ to the new experimentally established $\mathcal{B}(B\to {K }{\nu\nu})$, which is shown in Fig.~\ref{fig:corr1}. 
%%%%%%%%%%%%%%%%%%%%
%%%%%%%%%%%%%%%%%%%º%

Note, in particular, that we find the following relation between $B\rightarrow K^\ast\nu\bar{\nu}$ and $B\rightarrow K\nu\bar{\nu}$,%Note in particular that, under the assumption of new physics states being heavier than the electroweak symmetry breaking scale, we can find the following relation between $B\rightarrow K^\ast\nu\bar{\nu}$ and $B\rightarrow K\nu\bar{\nu}$, such that
\begin{equation}
\label{eq:eft-prediction}
    \frac{\mathcal{B}\left(B\rightarrow K^\ast\nu\bar{\nu}\right)}{\mathcal{B}\left(B\rightarrow K\nu\bar{\nu}\right)}\geq\frac{\mathcal{B}\left(B\rightarrow K^\ast\nu\bar{\nu}\right)}{\mathcal{B}\left(B\rightarrow K\nu\bar{\nu}\right)}\Bigg|_{\mathrm{SM}}\left(1-\frac{\eta_{K^\ast}}{4}\right),
\end{equation}
which is depicted by the hatched-dark blue region labelled as ``EFT'' in Fig.~\ref{fig:corr1} (see also \cite{Bause:2023mfe}).

By switching on either $\delta C_{L}^{\nu_i\nu_j}$ or $\delta C_{R}^{\nu_i\nu_j}$ and by considering all three diagonal and universal couplings to neutrino flavors, such that $\delta C_{L(R)}^{\nu_i\nu_j}=\delta_{ij}\delta C_{L(R)}$, we obtain that accommodating the measured $\mathcal{B}(B\to K{\nu\nu})$ to $2\sigma$ results in:
\begin{align}\label{eq:expCLR}
%&\delta C_{L} \in [-4.6, -3.5] \cup [ 16.1, 17.3] \,,\nonumber\\
&\delta C_{L,R} \in [-12.0, -3.5] \cup [ 16.1, 24.7]  \,.
\end{align}
These solutions become strongly restricted after imposing the experimental bound $\mathcal{B}(B\to {K^{\ast}}{\nu\nu})^\mathrm{exp} < 2.7 \times 10^{-5}$~\cite{Belle:2017oht}. In particular,  the large positive values of $\delta C_{R}$ are discarded, and the above ranges become: 
\begin{align}\label{eq:bounds1}
\delta C_{L} \neq 0 :  &\ \delta C_{L} \in  [-4.2, -3.5] \cup [ 16.9, 17.3],\quad \nonumber\\
                                  &  R_{\nu\nu}^{K^{\ast}} \in (2.4, 2.7) \,,\nonumber\\[0.5em]
\delta C_{R} \neq 0 : &\ \delta C_{R} \in [-12.0,-3.5],\quad \nonumber\\
                                 &R_{\nu\nu}^{K^{\ast}} \in (0.6, 2.1) \,, 
\end{align}
where we also give the resulting $\mathcal{B}(B\to {K^{\ast}}{\nu\nu})$ compatible with $\mathcal{B}(B\to K{\nu\nu})^\mathrm{exp}$ to $2\sigma$
 and the experimental bound on $\mathcal{B}(B\to {K^{\ast}}{\nu\nu})^\mathrm{exp}$. These allowed ranges are highlighted in Fig.~\ref{fig:corr1} with darker hues.
At this stage we consider the BSM contributions to be neutrino flavor universal, which is a relevant information when interpreting the values for $\delta C_{L,R}$. 
However, the correlation shown in Fig.~\ref{fig:corr1} and the ranges of $R_{\nu\nu}^{K^{\ast}}$
%bounds on $\mathcal{B}(B\to {K^{\ast}}{\nu\nu})$ 
given in Eq.~\eqref{eq:bounds1} remain 
as such regardless of whether we assume the NP to be lepton flavor universal or not.

\begin{figure}[t!]
\centering
\includegraphics[width=.999\linewidth]{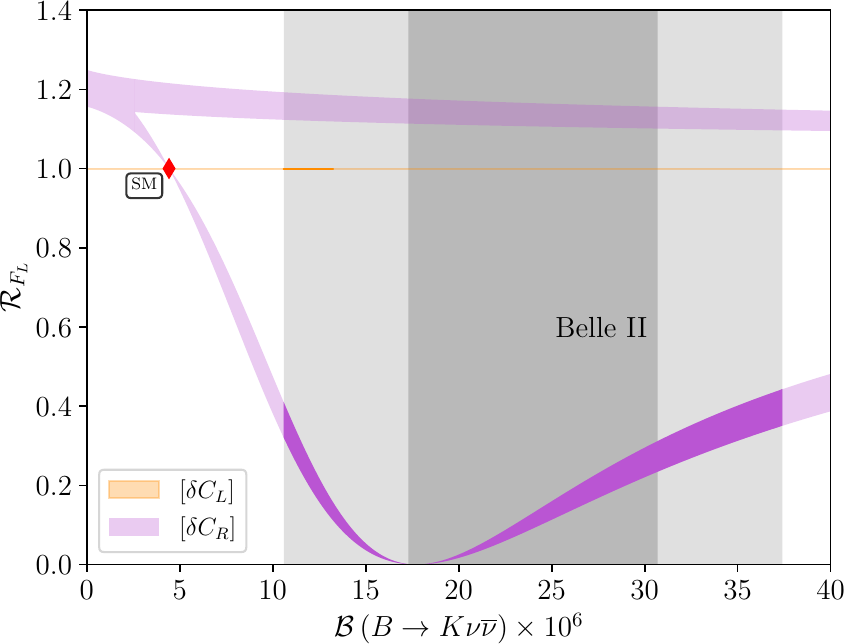}
\caption{\small \sl
Similar to Fig.~\ref{fig:corr1} we show how the fraction of $B\to {K^{\ast}}{\nu\nu}$ with longitudinally polarized $K^\ast$ responds to the variation of $\delta C_{L}$ or $\delta C_{R}$. Note that the yellow and purple curves correspond to those shown in Fig.~\ref{fig:corr1}, each associated with the allowed ranges of the Wilson coefficient given in Eq.~\ref{eq:bounds1}.}
%\GP{ The highlighted curves correspond to the highlighted regions of Fig.~\ref{fig:corr1}, for the values of the Wilson coefficient in Eq.~\ref{eq:bounds1}.}
\label{fig:corr2} 
\end{figure}

One can go a step further and predict the behavior of a fraction of the $B\to {K^{\ast}}{\nu\nu}$ corresponding to a specific polarization state of the outgoing $K^\ast$. For example, one can check how $F_L$,  the fraction of decay rate corresponding to the longitudinally polarized $K^\ast$ (cf. Refs.~\cite{Buras:2014fpa,Das:2017ebx}), responds to a non-zero $\delta C_{L}$ or $\delta C_{R}$.
$\mathcal{R}_{F_L} = F_L/F_L^\mathrm{SM}$ is presented in Fig.~\ref{fig:corr2}
%Such a plot is presented in Fig.~\ref{fig:corr2} in the notation $\mathcal{R}_{F_L} = F_L/F_L^\mathrm{SM}$ 
to better emphasize the fact that $F_L$ remains insensitive to $\delta C_{L}\neq 0$, whereas it becomes drastically depleted with respect to the SM value, $F_L^\mathrm{SM}=0.49(7)$, when $\delta C_{R}\neq 0$ is chosen such that it is consistent with both $\mathcal{B}(B\to K{\nu\nu})^\mathrm{exp}$ and the experimental bound on $\mathcal{B}(B\to K^\ast{\nu\nu})^\mathrm{exp}$, corresponding to negative values of $\delta C_R$ and the darker purple region in Fig.~\ref{fig:corr2}. In fact, we obtain that with $\delta C_{R}\neq 0$, the value of $F_L$ gets more than $50\%$ suppressed with respect to its SM value. From the plot in Fig.~\ref{fig:corr2} we read off:  
\begin{equation}
0 \leq \mathcal{R}_{F_L} \leq 0.44 \quad \Rightarrow \quad F_L \in [0,0.21]\,.
\end{equation}
This is a clear prediction that could be tested experimentally. Note, in particular, that $F_L$ has a much smaller theoretical uncertainty in the SM than the $\mathcal{B}(B\to K^\ast \nu \bar{\nu})$, since most of the form factor uncertainties cancel out in the ratio.

\section{SMEFT considerations}

When discussing the BSM scenarios in which the new degrees of freedom enter the stage well above the electroweak scale, it is convenient to work in the SM effective field theory (SMEFT), invariant under the full $SU(3)_c\times  SU(2)_L \times U(1)_Y$ gauge symmetry~\cite{Buchmuller:1985jz}. This allows one to relate the $b\to s\nu\bar{\nu}$ and $b\to s\ell\ell$ via $SU(2)_L$ gauge symmetry~\cite{Buras:2014fpa,Alonso:2014csa,Aebischer:2015fzz,Bause:2021cna,deGiorgi:2022vup,Rajeev:2021ntt}.
Of all the $d=6$ operators in 
\begin{equation}
\mathcal L^{(6)}_\mathrm{SMEFT} \supset  \sum_i \frac{\mathcal{C}_i}{\Lambda^2} {\mathcal O}_i\,,
\end{equation}
we select those relevant to our study, namely, 
\begin{align}
\begin{split}\label{eq:2fH}
\big{[}\mathcal{O}_{Hq}^{(1)}\big{]}_{kl} &= \big{(}\overline{Q}_k \gamma^\mu Q_l\big{)} \big{(}H^\dagger {\overleftrightarrow{D}}_\mu H \big{)}  \,, \\[0.4em]
\big{[}\mathcal{O}_{Hq}^{(3)}\big{]}_{kl} &= \big{(}\overline{Q}_k \tau^ I\gamma_\mu Q_l\big{)} \big{(}H^\dagger {\overleftrightarrow{D}}_\mu \tau^ I H \big{)}  \,,\\[0.4em]
\big{[}\mathcal{O}_{Hd}\big{]}_{kl} &=  \big{(}\overline{d}_{k R}\gamma_\mu d_{l R}\big{)}\big{(}H^\dagger {\overleftrightarrow{D}}_\mu   H \big{)}\,,\\[0.9em]
\end{split} \\
\begin{split}\label{eq:4f}
\big{[}\mathcal{O}_{lq}^{(1)}\big{]}_{ijkl} &= \big{(}\overline{L}_i\gamma^\mu L_j\big{)} \big{(}\overline{Q}_k \gamma_\mu Q_l\big{)}\,, \\[0.4em]
\big{[}\mathcal{O}_{lq}^{(3)}\big{]}_{ijkl} &=\big{(}\overline{L}_i\gamma^\mu \tau^ I L_j\big{)} \big{(}\overline{Q}_k \tau^ I\gamma_\mu Q_l\big{)}\,,\\[0.4em]
\big{[}\mathcal{O}_{ld}\big{]}_{ijkl} &=\big{(}\overline{L}_i\gamma^\mu  L_j\big{)} \big{(}\overline{d}_{k R}\gamma_\mu d_{l R}\big{)}\,,\\[0.4em]
\end{split}
\end{align}
where $ Q$ and $L$ denote the quark and lepton $SU(2)_L$ doublet, respectively, while $u, d, e$ stand for the quark and lepton weak singlets. 
In what follows, we will work in the flavor basis defined with diagonal down-type quark Yukawa matrix, i.e.~with the CKM matrix element 
in the upper component of $Q_i=[(V^\dagger\,u)_i\, ,\,d_i ]^T$.
Since we are focusing on the $b\to s$ processes we will fix $k=2$ and $l=3$, and discuss the two subsets of operators (\ref{eq:2fH},\ref{eq:4f}) separately. 
 
%%%%%%%%%%%%%%%%%%%%%%%%%%%%%%%%%%%%%%%%
\subsection{Quark bilinears and Higgs}\label{ssec:2fH}
%%%%%%%%%%%%%%%%%%%%%%%%%%%%%%%%%%%%%%%%

\begin{figure}[t!]
    \centering  
    \includegraphics[width=.999\linewidth]{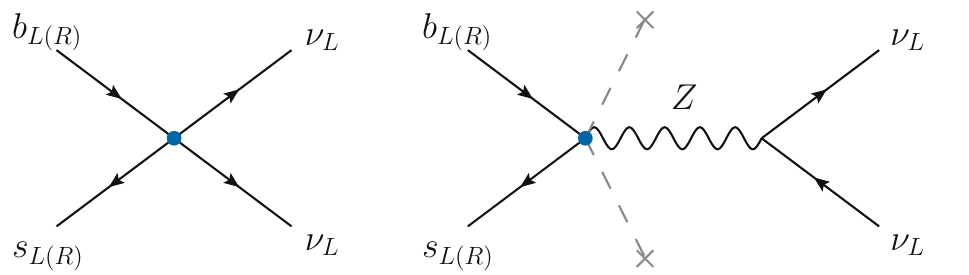}
    \caption{\small \sl Tree-level contributions of four-fermion (left panel) and Higgs-fermion (right panel) operators to the $b\to s\nu\bar{\nu}$ transition in the SMEFT.}
    \label{fig:diagrams}
\end{figure}

Firstly, we consider the operators with quark-bilinears and the Higgs current, as defined in Eq.~\eqref{eq:2fH}. These contribute to $B\to K\nu\bar{\nu}$ via a tree-level induced $Z$-coupling to the $(\bar s \gamma^\mu b)$ current, as depicted in the right panel of Fig.~\ref{fig:diagrams}. Clearly, these contributions would impact not only the $b\to s\nu_\ell\bar{\nu}_\ell$ transition, but also $b\to s\ell\ell$, with lepton-flavor universal contributions~\cite{Buras:2014fpa}, e.g.~the precisely determined $B_s \to \mu\mu$ branching ratio. Moreover, a double insertion of these coefficients will also have an effect in $B_s - \bar B_s$ mixing. 
Writing the $Z$-boson interaction Lagrangian with down-type quarks as
\begin{align}
\label{eq:Zll}
\hspace{-1.3em}\mathcal{L}_\mathrm{eff}^{Z} &= -\dfrac{g}{\cos \theta_W}\sum_{i,j} \bar{d}_i \gamma^\mu \left( g_{d_{L}}^{Z\,ij}\,P_L + g_{d_{R}}^{Z\,ij}\,P_R \right) d_j Z_\mu\,,
\end{align}
where we find
\begin{align}
\delta g_{d_L}^{Z\,ij} &= -\dfrac{v^2}{2\Lambda^2}\left\lbrace \big{[}\mathcal{C}_{Hq}^{(1)}\big{]}_{ij}+\big{[}\mathcal{C}_{Hq}^{(3)}\big{]}_{ij} \right\rbrace\,,\\[0.4em]
\delta g_{d_R}^{Z\,ij} &= -\dfrac{v^2}{2\Lambda^2}\big{[}\mathcal{C}_{Hd}\big{]}_{ij}\,,
\end{align}
with $g_\psi^{Z\, ij} = \delta_{ij}\, g_\psi^{Z} + \delta g_\psi^{Z\, ij}$, $g_{d_L}^{Z}=-1/2 +1/3 \sin^2\theta_W$ and $g_{d_R}^{Z}=+1/3 \sin^2\theta_W$, and $\theta_W$ the Weinberg angle. These relations can be used to match onto the relevant low-energy effective four-fermion operators, after integration of the SM vector bosons~\cite{Jenkins:2017jig}. In particular, one finds, for example, that the contribution from NP to $B_s\rightarrow\mu\mu$ depends on the SMEFT operators as $\delta C_{10}^{\ell_i\ell_i}\propto \big{[}\mathcal{C}^{(1)}_{Hq}\big{]}_{23}+\big{[}\mathcal{C}^{(3)}_{Hq}\big{]}_{23}-\big{[}\mathcal{C}_{Hd}\big{]}_{23}$.
Using the constraints from $\mathcal{B}(B_s \to \mu\mu)$~\cite{CMS:2022mgd} and $\Delta m_{B_s}$~\cite{Jenkins:2017dyc,Aebischer:2020dsw}, we determine the 2$\sigma$ confidence intervals for the coefficients $[\mathcal{C}_{Hq}^{(1)}]_{23}$, $[\mathcal{C}_{Hq}^{(3)}]_{23}$, and $[\mathcal{C}_{Hd}]_{23}$, switching on one at a time at $\Lambda=1$ TeV and accounting for the renormalization group evolution from $\Lambda$ down to $\mu=m_b$~\cite{Jenkins:2013zja}. The allowed intervals for the coefficients can then be translated onto intervals for the $B\to K \nu\bar\nu$ and $B\to K^\ast \nu\bar\nu$ branching fractions,
\begin{align}\label{eq:boundsQH}
\mathcal{C}^{(1)}_{Hq} \neq 0 :  &\ \mathcal{C}^{(1)}_{Hq}/\Lambda^2 \in  [-0.86, 0.45]\times 10^{-3}\,\mathrm{TeV}^{-2} ,\quad \nonumber\\
                                  &  R_{\nu\nu}^{K} \in (0.85, 1.08),\,R_{\nu\nu}^{K^{\ast}} \in (0.85, 1.08) \,,\nonumber\\[0.5em]
\mathcal{C}^{(3)}_{Hq} \neq 0 : &\ \mathcal{C}^{(3)}_{Hq}/\Lambda^2 \in [-0.78,0.41]\times 10^{-3}\,\mathrm{TeV}^{-2},\quad \nonumber\\
                                 &  R_{\nu\nu}^{K} \in (0.85, 1.08),\,R_{\nu\nu}^{K^{\ast}} \in (0.85, 1.08) \,,\nonumber\\[0.5em]
\mathcal{C}_{Hd} \neq 0 :  &\ \mathcal{C}_{Hd}/\Lambda^2 \in  [-0.45, 0.86]\times 10^{-3}\,\mathrm{TeV}^{-2} ,\quad \nonumber\\
                                  &  R_{\nu\nu}^{K} \in (0.92, 1.16),\,R_{\nu\nu}^{K^{\ast}} \in (0.90, 1.05) \,.
\end{align}

We find that the $\mathcal{B}(B^\pm\to K^\pm \nu\bar{\nu})$ value can only be enhanced by $\approx 20\%$, which is largely insufficient to accommodate the deviation shown in Belle-II data.

% \begin{table}[t]
% \centering
% \begin{tabular}{c|c|c}
% coeff. & $\rm{Br} (B^+\to K^+\nu\bar\nu) \,(\times 10^6)$ & $\rm{Br} (B^0\to K^{0\ast}\nu\bar\nu) \,(\times 10^6)$ \\ \hline\hline
%  & \multicolumn{2}{c}{$B_s\to\mu\mu$} \\ \hline
%  $[\mathcal{C}_{Hq}^{(1)}]_{23}$ & [4.01,5.06] & [8.23,10.4] \\
%  $[\mathcal{C}_{Hq}^{(3)}]_{23}$ & [4.01,5.06] & [8.23,10.4] \\
%  $[\mathcal{C}_{Hd}]_{23}$ & [4.32,5.42] & [8.71,10.2] \\ \hline
%  & \multicolumn{2}{c}{$B_s-\bar B_s$ mixing} \\ \hline
%  $[\mathcal{C}_{Hq}^{(1)}]_{23}$ & [4.01,4.68] & [8.23,9.63] \\
%  $[\mathcal{C}_{Hq}^{(3)}]_{23}$ & [4.01,4.68] & [8.23,9.63] \\
%  $[\mathcal{C}_{Hd}]_{23}$ & [4.68,5.42] & [8.71,9.63] \\
% \end{tabular}
% \caption{2$\sigma$ allowed intervals for $B\to K \nu\bar\nu$ and $B\to K^\ast \nu\bar\nu$ branching fractions taking into account limits from $B_s\to\mu\mu$ and $B_s$-$\bar B_s$ mixing. To be compared with the SM predictions $\rm{Br} (B^+\to K^+\nu\bar\nu)_{\rm SM} = (4.68 \pm 0.21)\times 10^{-6}$ and $\rm{Br} (B^0\to K^{0\ast}\nu\bar\nu)_{\rm SM} = (9.6 \pm 1.3)\times 10^{-6}$.}
% \label{tab:bounds}
% \end{table}

%%%%%%%%%%%%%%%%%%%%%%%%%%%%%%%%%%%%%%%%
\subsection{Four-fermion operators}\label{ssec:4f}
%%%%%%%%%%%%%%%%%%%%%%%%%%%%%%%%%%%%%%%%
We turn now our attention to the four-fermion operators defined in Eq.~\eqref{eq:4f}. Let us first explicitly rewrite $\mathcal L^{(6)}_\mathrm{SMEFT}$ using the operators given in Eq.~\eqref{eq:4f}. We have: 
\begin{align}\label{eq:LAGR}
&\mathcal L^{(6)}_\mathrm{SMEFT} \supset \frac{1}{\Lambda^2} \biggl\{  \left( \mathcal{C}_{lq}^{(1)} + \mathcal{C}_{lq}^{(3)} \right)_{ij}\,  ( \overline s_L \gamma^\mu b_L) ( \overline e_{L i} \gamma_\mu e_{L j}) 
\biggr. \nonumber\\
&\qquad+ \left( \mathcal{C}_{lq}^{(1)} - \mathcal{C}_{lq}^{(3)} \right)_{ij}\,   (\overline s_L \gamma^\mu b_L) ( \overline \nu_{L i} \gamma_\mu \nu_{L j})   \nonumber\\
&\qquad + 2\, V_{cs}\, \left[\mathcal{C}_{lq}^{(3)}\right]_{ij} \,  ( \overline c_L \gamma^\mu b_L) ( \overline e_{L i} \gamma_\mu \nu_{L j})  \nonumber\\
&\biggl. +  \left[\mathcal{C}_{ld}\right]_{ij}\, (\overline s_R \gamma^\mu b_R) \left[ ( \overline \nu_{L i} \gamma_\mu \nu_{L j}) + ( \overline e_{L i} \gamma_\mu e_{L j})\right] +\mathrm{h.c.}\biggr\}\,, 
\end{align}
where we have not written the charged-current operators that are CKM-suppressed. By comparing the above Lagrangian with the one given in Eq.~\eqref{eq:eft-bsnunu} it is easy to identify:
%%%%%%%%%%%%%
\begin{align}
\begin{split}
\label{eq:left-CL-bsnunu1}
\delta C_L^{\nu_i \nu_j} &= \dfrac{\pi}{\alpha_\mathrm{em} \lambda_t} \dfrac{v^2}{\Lambda^2} \left\lbrace\big{[}\mathcal{C}_{lq}^{(1)}\big{]}_{ij}-\big{[}\mathcal{C}_{lq}^{(3)}\big{]}_{ij}\right\rbrace \,, \\[0.4em]
\delta C_R^{\nu_i \nu_j} &= \dfrac{\pi}{\alpha_\mathrm{em} \lambda_t} \dfrac{v^2}{\Lambda^2} \big{[}\mathcal{C}_{l d}\big{]}_{ij}\,.
\end{split}
\end{align}
%%%%%%%%%%%%
As it was discussed in Ref.~\cite{Becirevic:2023aov}, several simple scenarios can be distinguished if we switch one of the above SMEFT operators at the time. For example, by allowing only $\mathcal{C}_{lq}^{(1)}\neq 0$ we get the simplest scenario with a $Z^\prime$ boson coupled only to the left-handed fermions. If, instead, the $Z^\prime$ is considered to be member of the weak triplet then one has only $\mathcal{C}_{lq}^{(3)}\neq 0$. Quite peculiar is the case with $\mathcal{C}_{lq}^{(3)}=\mathcal{C}_{lq}^{(1)}\neq 0$, which corresponds to the BSM scenario often encountered in the literature and involves a singlet vector leptoquark state of hypercharge $Y=2/3$. Similarly, a choice $\mathcal{C}_{lq}^{(1)}=3 \mathcal{C}_{lq}^{(3)}$ corresponds to the BSM model with a triplet of scalar leptoquarks with hypercharge $Y=1/3$~\cite{Buchmuller:1986zs}.

\section{Phenomenology}
So far we did not touch on the issue of lepton flavor. 
In this Section we discuss simple BSM scenarios and check whether or not one can build a simple scenario compatible with experimental data, and most importantly with 
$\mathcal{B}(B\to K{\nu\nu})^\mathrm{exp}$. In doing so we will separately treat the lepton flavor conserving case in which the couplings are diagonal ($i=j$), from the lepton flavor violating, i.e.~with non-diagonal couplings ($i\neq j$) being non-zero.

\subsection{New Physics coupling to muons only \label{ssec:mu}}

%%%%%%%%%%%%%%%%%%%%
\begin{figure}[t!]
\centering
\includegraphics[width=.999\linewidth]{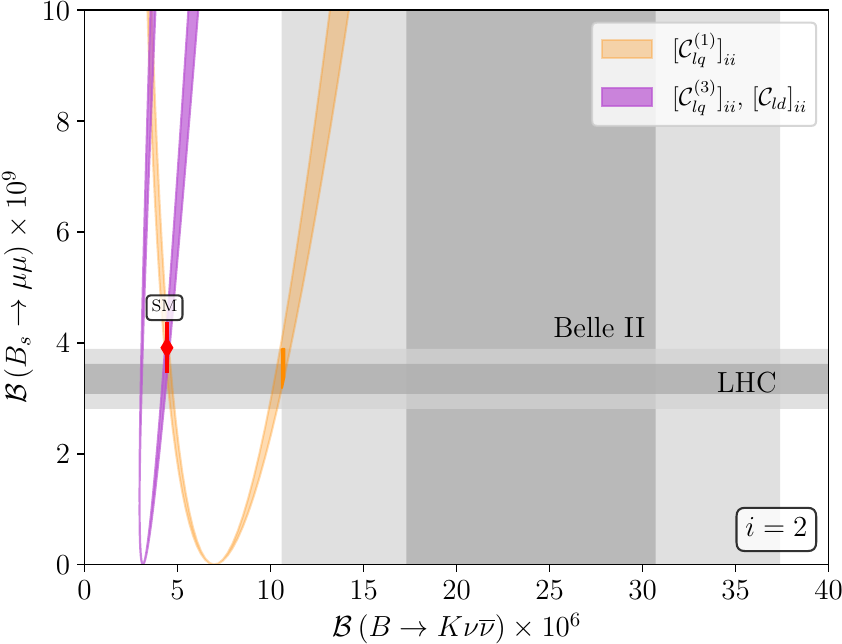}
\caption{\small \sl
Turning on the couplings to $\mu$'s only, also labeled as $i=2$, i.e. $\left[ \mathcal{C}_{lq}^{(1,3)}\right]_{22}\equiv  \mathcal{C}_{lq}^{(1,3)} \neq 0$ or $\left[\mathcal{C}_{ld}\right]_{22}\equiv \mathcal{C}_{ld}\neq 0$. Note that $\mathcal{C}_{lq}^{(3)}$ and $\mathcal{C}_{ld}$ lead to the same prediction (purple curve in the plot) which cannot be 
reconciled with $\mathcal{B}(B\to K{\nu\nu})^\mathrm{exp}$. Instead, by only allowing $\mathcal{C}_{lq}^{(1)}\neq 0$ (orange curve) one can get 
its small fraction to be consistent with both $\mathcal{B}(B\to K{\nu\nu})^\mathrm{exp}$ and $\mathcal{B}(B_s\to \mu\mu)^\mathrm{exp}$ to $2\sigma$, which is also highlighted in orange in the plot. The red point corresponds to the SM values (i.e. $\mathcal{C}_{lq}^{(1)}=\mathcal{C}_{lq}^{(3)} = \mathcal{C}_{ld} = 0$). }
\label{fig:corr3} 
\end{figure}
%%%%%%%%%%%%%%%%%%%%

We first attempt attributing the discrepancy between the measured $\mathcal{B}(B\to K{\nu\nu})$ and $\mathcal{B}(B\to K{\nu\nu})^\mathrm{SM}$ to the couplings to muon only, i.e. $i=j=2$. The best suited quantity to check the validity of such a solution is $\mathcal{B}(B_s\to {\mu\mu})$ for which we need to include the modification of the Wilson coefficient $C_{10}= C_{10}^\mathrm{SM}+ \delta C_{10}$, that in terms of $\mathcal{C}_{lq}^{(1,3)}$ and $\mathcal{C}_{ld}$ writes:
\begin{align}
\label{eq:C10-smeft}
 \delta C_{10}^{\ell_i\ell_i} &= \dfrac{\pi}{\alpha_\mathrm{em} \lambda_t} \dfrac{v^2}{\Lambda^2} \left\lbrace
  \big{[}\mathcal{C}_{ld}\big{]}_{ii} -
 \big{[}\mathcal{C}_{lq}^{(1)}\big{]}_{ii}-\big{[}\mathcal{C}_{lq}^{(3)}\big{]}_{ii}\right\rbrace\,, 
 \end{align}
In Fig.~\ref{fig:corr3} we show how $\mathcal{B}(B_s\to {\mu\mu})$ varies when switching  $\mathcal{C}_{lq}^{(1,3)}$ or $\mathcal{C}_{ld}$ one at the time. Clearly, 
any $\mathcal{C}_{lq}^{(3)}$ or $\mathcal{C}_{ld}$ cannot alone enhance $\mathcal{B}(B\to K{\nu\nu})^\mathrm{SM}$ in order to be consistent with experiment. 
The situation improves when 
$\mathcal{C}_{lq}^{(1)}\neq 0$ in which case one can reach the $2\sigma$ edge of $\mathcal{B}(B\to K{\nu\nu})^\mathrm{exp}$, while remaining consistent with $\mathcal{B}(B_s\to {\mu\mu})^\mathrm{exp}$ to $2\sigma$ as well. We find that the resulting allowed $\mathcal{C}_{lq}^{(1)}/\Lambda^2 \in [0.0129, 0.0131]\,\mathrm{TeV}^{-2}$, corresponding to the highlighted darker orange band in Fig.~\ref{fig:corr4}, and which translates into a very stringent bound: $2.16\leq \mathcal{B}(B\to {K^{\ast}}{\nu\nu}) \times 10^5 \leq 2.19$. 
This scenario, however, seems quite unlikely even though it would be consistent with the recent LHCb results regarding the lepton flavor universality $R_{K^{(\ast)}}$. In fact, consistency with $R_{K^{(\ast)}}$ would require $\mathcal{C}_{lq}^{(1)}/\Lambda^2 \in [0.0129, 0.0134]\,\mathrm{TeV}^{-2}$.

\subsection{New Physics coupling to two or three lepton species}

With respect to the previous case, the situation considerably improves if we turn on the couplings to both electrons and muons, in such a way that $\left[\mathcal{C}_{lq}^{(1)}\right]_{11}= \left[\mathcal{C}_{lq}^{(1)}\right]_{22}\equiv \mathcal{C}$. Quite obviously, in that case $R_{K^{(\ast)}}$ remains at its SM value and thus consistent with recent experimental analyses at LHCb. Compatibility with $\mathcal{B}(B\to K{\nu\nu})^\mathrm{exp}$ is so improved that even a small $1\sigma$ overlap with $\mathcal{B}(B\to K{\nu\nu})^\mathrm{exp}$ can be reached. 
 %%%%%%%%%%%%%%%%%%%%
\begin{figure}[t!]
\centering
\includegraphics[width=.999\linewidth]{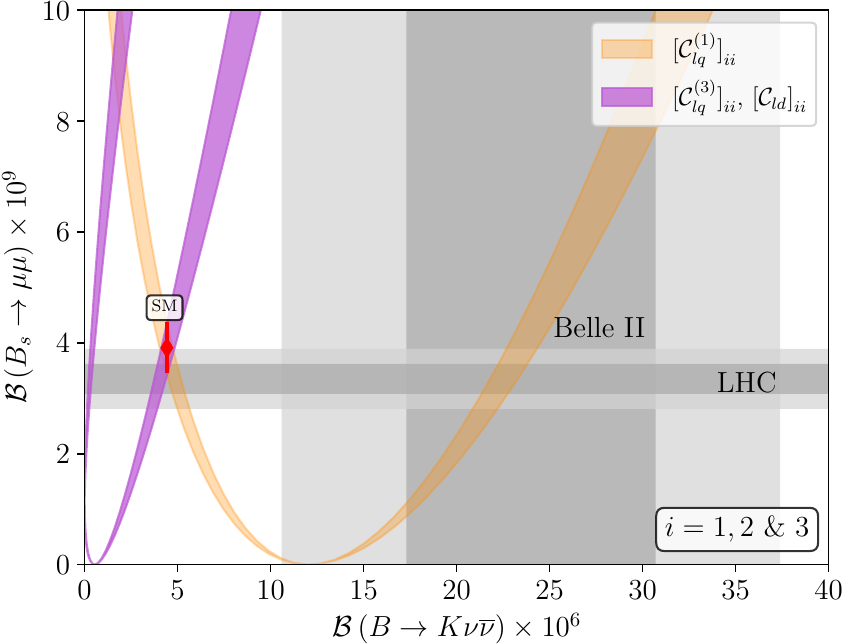}
\caption{\small \sl
Similar to Fig.~\ref{fig:corr3}, except that in this case we turn on all of the lepton species with $\left[\mathcal{C}\right]_{11}=\left[\mathcal{C}\right]_{22}= \left[\mathcal{C}\right]_{33}$ where $\mathcal{C}$ stands for either $\mathcal{C}_{lq}^{(1)}$, $\mathcal{C}_{lq}^{(3)}$ or $\mathcal{C}_{ld}$. }
\label{fig:corr4} 
\end{figure}
%%%%%%%%%%%%%%%%%%%%
To go deep into the $1\sigma$ region of both $\mathcal{B}(B\to K{\nu\nu})^\mathrm{exp}$ and $\mathcal{B}(B_s \to \mu\mu)$ one can take the flavor universal situation, and assume $\left[\mathcal{C}_{lq}^{(1)}\right]_{11}=\left[\mathcal{C}_{lq}^{(1)}\right]_{22}= \left[\mathcal{C}_{lq}^{(1)}\right]_{33}\equiv \mathcal{C}$, which is what we plot in Fig.~\ref{fig:corr4}. 
 Like before, $R_{K^{(\ast)}}$ remains unchanged with respect to its SM value, and overall consistency with the $b\to s$ data is achieved. However, as a result of the severe 
constraint arising from $\mathcal{B}(B_s\to {\mu\mu})^\mathrm{exp}$, in both cases we obtain 
$ \mathcal{C}/\Lambda^2 \in [0.012, 0.013]\,\mathrm{TeV}^{-2}$, which then results in either $R^{K^\ast}_{\nu\nu} \in (3.6,3.9)_{e,\mu}$ or $R^{K^\ast}_{\nu\nu} \in (4.8,5.3)_{e,\mu,\tau}$. Since both these intervals are larger than $R^{K^\ast\, \mathrm{(exp)}}_{\nu\nu}< 2.7$, all three scenarios discussed so far cannot meet the experimental constraints.

Another important shortcoming of the two scenarios discussed here, is that the contribution to ratios $R_D^{(\ast)}= \mathcal{B}(B\to D^{(\ast)} \tau\bar{\nu})/
\mathcal{B}(B\to D^{(\ast)} l \bar{\nu})$, where $l\in (e,\mu)$, is absent, since this needs to go through the triplet operator $\mathcal{C}_{lq}^{(3)}$ (cf. Eq.~\eqref{eq:LAGR}). 
We remind the reader that the most recent averages of experimental values for $R_{D^{(\ast)}}$~\cite{HeavyFlavorAveragingGroup:2022wzx}, 
\begin{equation}
 R_D = 0.257(29) , \quad  R_{D^{\ast}} = 0.284(12)\,
 \end{equation}
are larger than predicted in the SM~\cite{FlavourLatticeAveragingGroupFLAG:2021npn,MILC:2015uhg,FermilabLattice:2021cdg}, leading to $R_D/R_D^\mathrm{SM}= 1.19(10)$ and $R_{D^{\ast}}/R_{D^{\ast}}^\mathrm{SM}= 1.15(5)$, which can be averaged for our purpose to: 
\begin{equation}\label{eq:RD-s}
R_{D^{(\ast)}}^\mathrm{exp}/R_{D^{\ast}}^\mathrm{SM} = 1.16(5)\,.
 \end{equation}
From Figure \ref{fig:corr3} one can clearly see that the constraints from $B_s\to\mu\mu$ cannot be respected in the case $\left[\mathcal{C}_{lq}^{(3)}\right]_{11}=\left[\mathcal{C}_{lq}^{(3)}\right]_{22}= \left[\mathcal{C}_{lq}^{(3)}\right]_{33}$. Moreover, this lepton-flavor universal scenario predicts $R_{D^{(\ast)}}=R_{D^{(\ast)}}^\mathrm{SM}$ which disagrees with Eq.~\eqref{eq:RD-s}. Also the case $\left[\mathcal{C}_{lq}^{(3)}\right]_{11}=\left[\mathcal{C}_{lq}^{(3)}\right]_{22}$is in disagreement with data, since it predicts $R_{D^{(\ast)}}<R_{D^{(\ast)}}^\mathrm{SM}$ in correlation with an enhanced $\mathcal{B}(B\to K{\nu\nu})$.

One concludes that modification of the couplings to $\tau$'s is what is desired in order to be consistent with experimental data. More specifically one needs $\left[\mathcal{C}_{lq}^{(3)}\right]_{33}\neq 0$.

\subsection{New Physics coupling to taus only\label{sec:tau-only}}

\begin{figure}[t!]
\centering
\includegraphics[width=.999\linewidth]{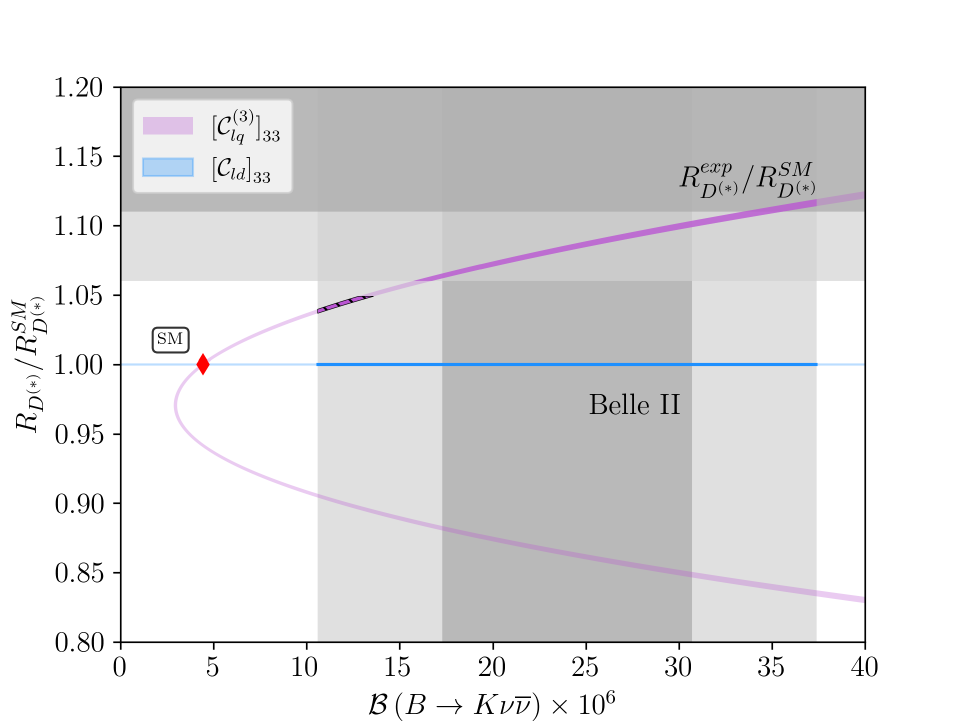}
\caption{\small \sl
The curve shows $R_{D^{(\ast)}}/R_{D^{\ast}}^\mathrm{SM}$ as a function of $\mathcal{B}(B\to K{\nu\nu})$ varying $\bigl[\mathcal{C}_{lq}^{(3)}\bigr]_{33}$ (purple) or $\bigl[\mathcal{C}_{ld}\bigr]_{33}$ (blue). The highlighted purple region correspond to $\bigl[\mathcal{C}_{lq}^{(3)}\bigr]_{33}$ allowed by $\mathcal{B}(B\to K{\nu\nu})^\mathrm{exp}$ which becomes restricted to the hatched region once the condition $R^{K^\ast\, \mathrm{(exp)}}_{\nu\nu}< 2.7$ is imposed. That range is smaller than $R_{D^{(\ast)}}^\mathrm{exp}/R_{D^{\ast}}^\mathrm{SM}$ to $2\sigma$ (horizontal gray area).
%from which we see that only one of the two solutions consistent with $2\sigma$ range of $\mathcal{B}(B\to K{\nu\nu})^\mathrm{exp}$ (vertical gray area) is consistent with  $R_{D^{(\ast)}}^\mathrm{exp}/R_{D^{\ast}}^\mathrm{SM}$ (horizontal gray area), which is highlighted in the plot.
The red point corresponds to the SM.}
\label{fig:corr5} 
\end{figure}

Finding $\mathcal{C}_{lq}^{(3)}\equiv \left[\mathcal{C}_{lq}^{(3)}\right]_{33}\neq 0$ such that $\mathcal{B}(B\to K{\nu\nu})$ and $R_{D^{(\ast)}}$ are simultaneously consistent with their experimental values is possible as we avoid the constraint coming from $\mathcal{B}(B_s\to {\mu\mu})^\mathrm{exp}$. 
We note that in this situation 
\begin{equation}\label{eq:RD-s-expr}
\frac{R_{D^{(\ast)}}  }{ R_{D^{(\ast)}}^\mathrm{SM}} = \left( 1 - \frac{v^2 }{\Lambda^2} \, \frac{V_{cs}}{V_{cb}}\, \mathcal{C}_{lq}^{(3)}\right)^2\,,
 \end{equation}
from which we see that $\mathcal{C}_{lq}^{(3)}\neq 0$ is significantly enhanced by $V_{cs}/V_{cb} = 24(1)$.
The solution is shown in Fig.~\ref{fig:corr5}. 
%%%%%%%%%%%%%%%%%%%%
%%%%%%%%%%%%%%%%%%%%
We find that the acceptable $\mathcal{C}^{(3)}_{lq}$ falls in:
\begin{equation}\label{eq:222}
\mathcal{C}^{(3)}_{lq} \in [-0.039,-0.019]\,\mathrm{TeV}^{-2}.
\end{equation}
As can be seen from \eqref{eq:RD-s-expr}, and in view of Eq.~\eqref{eq:RD-s}, only negative $\mathcal{C}^{(3)}_{lq}$ values will lead to $R_{D^{(\ast)}}^\mathrm{exp}/R_{D^{\ast}}^\mathrm{SM} >1$, and we obtain $R_{D^{(\ast)}}/R_{D^{\ast}}^\mathrm{SM}\in [1.06,1.12]$,
also highlighted in Fig.~\ref{fig:corr5}.
%\salva{, corresponding to the darker}\GP{(probably shaded is better? darker may be misleading)}\salva{ purple band in Fig.~\ref{fig:corr5}}. 

However, the $\mathcal{C}^{(3)}_{lq}$ range allowed by $R^{K^\ast\, \mathrm{(exp)}}_{\nu\nu}< 2.7$ and $\mathcal{B}(B\to K{\nu\nu})^\mathrm{exp}$ is $\mathcal{C}^{(3)}_{lq}/\Lambda^2 \in [-0.015,-0.013]\cup[0.033,0.036]\,\mathrm{TeV}^{-2}$, incompatible with Eq.~\eqref{eq:222}. Nonetheless, including other operators contributing to $b\rightarrow c\tau\nu$, but not to $b\rightarrow s\nu\bar{\nu}$, one can find scenarios allowing us to simultaneously explain the deviations in $R_K^{\nu\nu}$ and $R_{D^{(\ast)}}$. Example of such scenarios are $\mathcal{C}_{lq}^{(1)}=\mathcal{C}_{lq}^{(3)}$~\cite{Buttazzo:2017ixm}, or combinations of scalar $\mathcal{C}_{lequ}^{(1)}$ and tensor $\mathcal{C}_{lequ}^{(3)}$, e.g.~Ref.~\cite{Feruglio:2018fxo}.

As it is well known, accommodating $R_{D^{(\ast)}}^{\rm{(exp)}}$ results in an increase of decay rates to a pair of $\tau$-leptons in the final state~\cite{Capdevila:2017iqn}. Indeed we find that the scenarios compatible with $R_{D^{(\ast)}}^{\rm{(exp)}}$ and $\mathcal{B}(B\to K{\nu\nu})^{\rm{(exp)}}$, and consistent with the $\mathcal{C}^{(3)}_{lq} $ values given in Eq.~\eqref{eq:222} yield:
\begin{align}
\label{eq:pred-tautau-1}
\frac{ \mathcal{B}(B_s\to \tau\tau ) }{ \mathcal{B}(B_s\to \tau\tau )^\mathrm{SM}} & \in [15,46], \nonumber\\[0.5em]
\frac{ \mathcal{B}(B \to K \tau\tau ) }{ \mathcal{B}(B\to K \tau\tau )^\mathrm{SM}}
&= \frac{ \mathcal{B}(B \to K^\ast \tau\tau ) }{ \mathcal{B}(B\to K^\ast \tau\tau )^\mathrm{SM}}  
 \in [15, 49] \,.
 \end{align}
 
If we drop the requirement of compatibility with $R_{D^{(\ast)}}^{\rm{(exp)}}$ but instead insist on respecting the bound on $R_{\nu\nu}^{K^\ast(\mathrm{exp})}$, we find the following values,~\footnote{Note that we neglect the range for which $R_{D^{(\ast)}}^\mathrm{exp}/R_{D^{\ast}}^\mathrm{SM} <1$.}
\begin{align}
\label{eq:pred-tautau-2}
\frac{ \mathcal{B}(B_s\to \tau\tau ) }{ \mathcal{B}(B_s\to \tau\tau )^\mathrm{SM}} & \in [9,10], \nonumber\\[0.5em]
\frac{ \mathcal{B}(B \to K \tau\tau ) }{ \mathcal{B}(B\to K \tau\tau )^\mathrm{SM}}
&= \frac{ \mathcal{B}(B \to K^\ast \tau\tau ) }{ \mathcal{B}(B\to K^\ast \tau\tau )^\mathrm{SM}}  
 \in [9,11]\,.
 \end{align}

\noindent Note, in particular, that we only consider $\mathcal{O}_{lq}^{(3)}$ in the predictions shown in Eq.~\eqref{eq:pred-tautau-1}--\eqref{eq:pred-tautau-2}. However, the connection between the $b\to s\nu_\tau\bar{\nu}_\tau$ and $b\to s\tau\tau$ transitions is more general, as the operators $\mathcal{O}_{lq}^{(1)}$  and $\mathcal{O}_{ld}$ also contribute to both transitions. Therefore, if $\mathcal{B}(B\to K\nu\nu)$ is confirmed to be considerably larger than the SM value, one should expect a sizable deviation from the SM in $\mathcal{B}(B_s\to \tau\tau)$ and $\mathcal{B}(B\to K^{(\ast)}\tau\tau)$ as well. The only exception to this conclusion is the scenario where $\mathcal{C}_{lq}^{(1)}=-\mathcal{C}_{lq}^{(3)}$, which only affects neutral currents with neutrinos, cf.~Eq.~\eqref{eq:LAGR}.

\subsection{Lepton flavor violating case}

We can now assume that the difference between $\mathcal{B}(B\to K{\nu\nu})^\mathrm{exp}$ and its SM prediction can be described by turning on the off-diagonal couplings to lepton flavors ($i\neq j$). This can be done with a number of various assumptions. Here, as an example, we turn either  $\bigl[\mathcal{C}_{lq}^{(3)}\bigr]_{ij}$, or $\bigl[\mathcal{C}_{lq}^{(1)}\bigr]_{ij}$, or $\bigl[\mathcal{C}_{ld} \bigr]_{ij}$ at the time and assume the couplings to $\mu$ and $\tau$ to be non-zero by taking $\mathcal{C}_{32}=\mathcal{C}_{23}$. As a result we get that the $2\sigma$ compatibility with $\mathcal{B}(B\to K{\nu\nu})^\mathrm{exp}$ is achieved for 
$\mathcal{C}_{32}/\Lambda^2=\mathcal{C}_{23}/\Lambda^2\in [-0.034,-0.015]\cup[0.015,0.034]\,\mathrm{TeV}^{-2}$, which, after imposing $R^{K^\ast\, \mathrm{(exp)}}_{\nu\nu}< 2.7$, reduces to 
$\mathcal{C}_{32}/\Lambda^2=\mathcal{C}_{23}/\Lambda^2\in [-0.017,-0.015]\cup[0.015,0.017]\,\mathrm{TeV}^{-2}$. As a consequence the decay rates of the corresponding lepton flavor violating modes will be significant. 
By using the expressions derived in Ref.~\cite{Becirevic:2016zri} we obtain: 
\begin{align}
&\mathcal{B}(B\to K \mu\tau) \in [2, 3] \times 10^{-6},
\end{align}
not far from but consistent with the experimental bound~\cite{LHCb:2022wrs,ParticleDataGroup:2022pth}, $\mathcal{B}\left(B\to K\mu\tau\right)^{\mathrm{exp}}<4.8\times 10^{-5}$ at $90\%$~C.L.

\subsection{Expectations from concrete models}
So far we have been agnostic about the possible origin of these effective operators, not focusing on any concrete model~\cite{Angelescu:2018tyl,Greljo:2015mma,Buchmuller:1986zs}. It is worth mentioning that a BSM model with a weak triplet of scalar leptoquarks $S_3$ cannot be made consistent with all the data discussed here. The reason is well known, namely it cannot give a significant increase to $R_{D^{(\ast)}}$~\cite{Angelescu:2018tyl,Greljo:2015mma}. The $U_1$ vector leptoquark, in a simplified model approach (involving couplings to the left-handed operators only), cannot be made consistent with data. This is so because at tree-level $\mathcal{C}_{lq}^{(1)}=\mathcal{C}_{lq}^{(3)}$, or $\delta C_L =0$, which is excluded by the new Belle~II result for $\mathcal{B}(B\to K {\nu\nu})^\mathrm{exp}$. It is important to stress, however, that the $U_1$ being a massive vector necessarily calls for a UV completion, which typically requires also the inclusion of additional fermionic degrees of freedom. The full theory of the $U_1$ has been extensively studied, and in particular it has been shown that the relation $\mathcal{C}_{lq}^{(1)}=\mathcal{C}_{lq}^{(3)}$ is broken at one-loop, leading to an increase of $\mathcal{B}(B\to K {\nu\nu})$ of up to $\approx$ 50\% with respect to the Standard Model, given the current values of $R_{D^{(\ast)}}$~\cite{Fuentes-Martin:2019ign, Fuentes-Martin:2020luw}. 
Another possibility is to take the couplings to $\tau$'s verifying $\mathcal{C}_{lq}^{(1)}=-\mathcal{C}_{lq}^{(3)}$, which is valid in a scenario with the so called $S_1$ leptoquark. In that case one gets $R_{D^{(\ast)}}/R_{D^{\ast}}^\mathrm{SM} \approx 1.02$, thus much smaller than in the case discussed in Sec.~\ref{sec:tau-only}. 
The way out would then be to turn on the right handed couplings~\cite{Angelescu:2018tyl}. Similarly, in the case of the so called $\widetilde{R}_2$ leptoquarks one needs to turn on the right-handed couplings to quarks (and left-handed to $\tau$), which in the SMEFT language means that $\mathcal{C}_{ld}\neq 0$~\cite{Becirevic:2016yqi}. It thus can explain the new experimental result $\mathcal{B}(B\to K {\nu\nu})^\mathrm{exp}$. Moreover, by combining the constraints $\mathcal{B}(B\to K {\nu\nu})^\mathrm{exp}$ and $[\Delta m_s]^\mathrm{exp}/[\Delta m_s]^\mathrm{SM}$ we get the bound $m_{\widetilde R_2} \lesssim 3$~TeV and $m_{S_1} \lesssim 4.5$~TeV. This scenario, however, fails to explain $R_{D^{(\ast)}}^\mathrm{exp}/R_{D^{\ast}}^\mathrm{SM}$. Another similar model is the one with a $Z^\prime$ boson coupled to left-handed $\tau$ and $\bar s\gamma_\mu P_R b$, again giving rise to $\mathcal{C}_{ld}\neq 0$~\cite{Becirevic:2015asa}. While it can accommodate all of the $b\to s$ constraints, it does not contribute to $b\to c\tau \nu$. 

\subsection{Summary}

In this letter, we discuss the possible consequences on $\mathcal{B}(B\to K^\ast {\nu\nu})$ due to the recent Belle~II measurement of $\mathcal{B}(B\to K  {\nu\nu})$ which was found to be nearly $3\sigma$ larger than predicted. We find that the values  $\mathcal{B}(B\to K^\ast {\nu\nu})\lesssim 1.8 \times \mathcal{B}(B\to K^\ast {\nu\nu})^\mathrm{SM}$ could only occur if the coupling to the right-handed operator $\bar b_R \gamma_\mu s_R$ is non-zero, while larger branching fractions would be consistent with couplings to either left-handed or right-handed operators (or a combination of both). 

By relying on SMEFT we show that the increase in $\mathcal{B}(B\to K {\nu\nu})$ cannot be described by switching on the couplings to one or more lepton flavors if that choice comprises the coupling to muons. The reason is that $\mathcal{B}(B_s\to \mu\mu)^\mathrm{exp}$ provides a very stringent constraint on the values of the Wilson coefficient $\mathcal{C}_{lq}^{(1)}$, which is the only one that can provide the desired enhancement to $\mathcal{B}(B\to K^{(\ast )} {\nu\nu})$. In the SMEFT Lagrangian the Wilson coefficients relevant to $b\to s \ell\ell$ and $b\to s \nu\nu$ decay modes are also related to the semileptonic $b\to c\ell \nu$, so that one should make sure that the resulting $R_{D^{(\ast)}} > R_{D^{\ast}}^\mathrm{SM}$, as suggested by experiment. For that it is necessary that $\mathcal{C}_{lq}^{(3)}\neq 0$ or to have the scalar and tensor operators that are not related to $b\to s \ell\ell$ and $b\to s \nu\nu$~\cite{Angelescu:2018tyl}. We find that a significant increase in $R_{D^{(\ast)}}/R_{D^{\ast}}^\mathrm{SM}$ can be achieved if we allow only the coupling to $\tau$ and not to other species. In this way one can find the agreement with all of the above-mentioned experimental constraints, including $R_{K^{(\ast )}}$, except that $R_{D^{(\ast)}}/R_{D^{\ast}}^\mathrm{SM} \approx 1.05$.

We also considered a possibility with the off-diagonal couplings to lepton species. Accommodating $\mathcal{B}(B\to K  {\nu\nu})^\mathrm{exp}$ in this setting implies a large  $\mathcal{B}(B\to K  \mu\tau)$, not too far from the current experimental bound. 

We  need to emphasize that all the couplings discussed in this paper are well within the ranges allowed by the experimental data collected in the region of high-$p_T$ tails of the differential distribution of $pp\to \ell\ell$ (+ soft jets), cf. Ref.~\cite{Allwicher:2022gkm}.

One final comment regards the experimental value used in our analysis, which is the new measurement of $\mathcal{B}(B\to K\nu\nu)$ only. In \cite{Belle-II:2023esi}, also an average with the previous bound from Belle-II was provided \cite{Belle:2017oht}, giving $\mathcal{B}(B \to K \nu\nu)^{\exp} = 1.4(4) \times 10^{-5}$. How considering this average affects our conclusions can be mostly read off the same plots we presented, the main effect being a higher compatibility with the bound on $R_\nu^{K^\ast}$ in the left-handed scenario, and an increased (although still below the SM) range for $F_L$. Similarly, the $\tau$-only coupling scenario would call for a larger right-handed contribution in order to enhance the contributions to $R_D^{(\ast)}$.

%%%%%% Bibliography %%%%%%
%%%%%%%%%%%%%%%%%%%%
\section*{Acknowledgments} 
This project has received funding /support from the European Union’s Horizon 2020 research and innovation programme under the Marie Skłodowska-Curie grant agreement No 860881-HIDDeN and No 101086085-ASYMMETRY. The work of LA is supported by the Swiss National Science Foundation (SNF) under contract 200020-204428. LA is also grateful to the Mainz Institute for Theoretical Physics (MITP) of the Cluster of Excellence PRISMA+ (Project ID 39083149), for its hospitality and its partial support during the completion of this work.

\section*{Note added}

While this paper was in writing the results of studies similar to our's were released in Refs.~\cite{Bause:2023mfe,Athron:2023hmz}. In particular the authors of Ref.~\cite{Bause:2023mfe} interpret the new Belle~II results in terms of the SMEFT operators reaching the conclusions which agree with ours for the most part.

%%%%%%%%%%%%%%%%%%%%%
%%%%%%%%%%%%%%%%%%%%%

%%%%%%%%%%%%%%%%%%%%%
%%%%%%%%%%%%%%%%%%%%%

%%%%%%%%%%%%%%%%%%%%%%%%%%
\end{document}